\documentclass[superscriptaddress, twocolumn, aps, pra]{revtex4-2}
\usepackage{amsfonts,amssymb,amsmath,amsthm}
\usepackage[]{graphics,graphicx,epsfig}
\usepackage{epstopdf}
\usepackage{graphicx}
\usepackage{natbib}
\usepackage{color}
\usepackage{xcolor}
\newcommand{\ket}[1]{|{#1}\rangle}
\newcommand{\bra}[1]{\langle{#1}|}
\usepackage{array}
\newcolumntype{P}[1]{>{\centering\arraybackslash}p{#1}}

\begin{document}

\title{Permutation asymmetry unlocks  emergent advantage in randomized Bell tests}

\author{Wies{\l}aw Laskowski}
\affiliation{Institute of Theoretical Physics and Astrophysics, University of Gda\'nsk, 80-308 Gda\'nsk, Poland}

\author{Tam{\'a}s V{\'e}rtesi}
\affiliation{HUN-REN Institute for Nuclear Research, P.O. Box 51, H-4001 Debrecen, Hungary}

\author{Jan W{\'o}jcik}
\affiliation{Institute of Theoretical Physics and Astrophysics, University of Gda\'nsk, 80-308 Gda\'nsk, Poland}

\begin{abstract}
All maximally entangled two-qubit states violate local realism with the same probability under uniformly random projective measurements, yet they need not behave identically in randomized Bell tests. We show that when measurement settings are exchanged between the parties in sequential Bell experiments, permutation symmetry of the shared state determines the statistical relation between the two realizations. Permutationally invariant states yield identical nonlocality outcomes in both experiments, whereas asymmetric states can violate local realism in one realization but not in the other. This distinction leads to two operational consequences. First, it enables the detection of correlations between the measurement choices of Alice and Bob through the joint violation statistics. Second, in Bell tests with finite measurement pools, asymmetric maximally entangled states can significantly increase the probability of observing nonlocality without requiring additional resources. Our results identify permutation asymmetry as a useful feature in randomized Bell experiments and highlight a new role of symmetry in quantum nonlocality.
\end{abstract}

\maketitle

\section{Introduction} 

Bell nonlocality provides a fundamental distinction between quantum and
classical descriptions of physical systems, demonstrating the incompatibility
of quantum correlations with local hidden-variable
models~\cite{PhysicsPhysiqueFizika.1.195}. Since the formulation of the
Clauser-Horne-Shimony-Holt (CHSH) inequality~\cite{PhysRevLett.23.880}, violations of local
realism have become a key resource in quantum information processing, forming
the basis of device-independent protocols such as quantum key distribution,
randomness expansion, and self-testing~\cite{RevModPhys.86.419}.

Standard Bell scenarios treat the two observers, Alice and Bob, symmetrically.
It is now well understood, however, that quantum correlations can exhibit
various forms of directional asymmetry. A prominent example is quantum
steering~\cite{RevModPhys.92.015001}, where the roles of the parties are inherently inequivalent,
giving rise to phenomena such as one-way steering~\cite{PhysRevLett.112.200402}.
These results establish that nonclassical correlations depend not only on the
underlying quantum state, but also on the assignment of measurement roles to
the observers.

Motivated by this interplay between correlations and measurement roles, we
investigate Bell nonlocality under random measurements in a setting where the
measurement configurations are physically exchanged between the parties.
Although this exchange does not introduce an operational asymmetry of the type
encountered in steering, it probes, in a related spirit, how nonclassical
correlations respond to a swap of roles. Specifically, we analyze how
exchanging measurement settings between Alice and Bob affects the probability
of observing nonlocal correlations in sequential experiments. We show that even
among maximally entangled states, the presence or absence of permutation
symmetry leads to qualitatively different statistical correlations between the
two experimental realizations, with direct consequences for the efficiency of
randomized Bell tests.

More generally, our findings illustrate that permutation asymmetry, while irrelevant for the single-test probability of Bell violation, can acquire operational significance when multiple randomized Bell experiments are considered jointly, echoing broader studies of asymmetry as a useful quantum feature~\cite{Bartlett2007,Gour2008}. Note that throughout this work Bell violation is defined with respect to the full local Bell polytope, rather than a single Bell expression that may be asymmetric under party exchange.

\section{ Standard nonlocality randomized measurement}

\label{sec:model}

We begin by outlining the standard scheme for demonstrating nonlocality via
randomized measurements. In two spatially separated laboratories, Alice and Bob
independently choose between two dichotomic observables, $A_1, A_2$ and
$B_1, B_2$, respectively, thereby defining a standard Bell experiment.
Correlations that cannot be explained by any local realistic model are witnessed
by violations of Bell inequalities. For the two-setting, two-outcome scenario
considered here, the relevant figure of merit is the CHSH expression
\begin{equation}
    I = \langle A_1 B_1 \rangle + \langle A_1 B_2 \rangle
      + \langle A_2 B_1 \rangle - \langle A_2 B_2 \rangle,
\end{equation}
which satisfies $|I| \leq 2$ for any local realistic model, while quantum
mechanics permits $|I| = 2\sqrt{2}$ for suitable measurement settings.

Equivalently, local realism holds if and only if there exists a joint
probability distribution reproducing all experimentally accessible
marginals~\cite{PhysRevLett.48.291}. Rather than focusing on a specific inequality, we adopt a
probabilistic viewpoint and treat the non-existence of such a joint probability distribution as an event
in the space of measurement choices, verifiable via linear programming (e.g.~\cite{PhysRevLett.85.4418}).

Within this framework, we define $P_V$ as the probability that randomly chosen
observables $A_1, A_2$ and $B_1, B_2$ yield correlations incompatible with any
local realistic model. For the two-qubit singlet state $|\psi^-\rangle$, with
measurement directions sampled uniformly over the Bloch sphere, this
probability is analytically known to be $P_V = 2(\pi - 3)$~\cite{PhysRevLett.104.050401}.

\section{Modified scenario: swapped measurement settings}

\label{sec:scenario}
We now generalize this scheme by allowing Alice and Bob to exchange their
measurement settings between two successive experiments. In the first
experiment, $S_1$, Alice measures $A_1$ and $A_2$ while Bob measures $B_1$
and $B_2$. In the second experiment, $S_2$, the roles are exchanged: Alice
measures $B_1$ and $B_2$ while Bob measures $A_1$ and $A_2$.

This leads naturally to four probabilities of interest:
\begin{itemize}
    \item $P_{S_1}$: probability that the correlations observed in $S_1$ are
          incompatible with a local realistic model;
    \item $P_{S_2}$: probability that the correlations observed in $S_2$ are
          incompatible with a local realistic model;
    \item $P_{\mathrm{and}} \equiv P_{S_1 \wedge S_2}$: probability that both
          experiments simultaneously yield nonlocal correlations;
    \item $P_{\mathrm{or}} \equiv P_{S_1 \vee S_2}$: probability that at least
          one experiment yields nonlocal correlations.
\end{itemize}
These four quantities are related by the inclusion-exclusion identity
\begin{equation}
P_{\text{or}} = P_{S_1} + P_{S_2} - P_{\text{and}},
\label{ee}
\end{equation}
which will play a central role in the operational analysis of finite measurement pools.

\section{Violation statistics under swapped measurements}

\label{sec:results}
Because the observables are drawn from a uniform random distribution, ensemble
averaging dictates $P_{S_1} = P_{S_2}$. Furthermore, for all maximally
entangled two-qubit states, the single-experiment violation probability is
$P_{S} = 2(\pi - 3)$. Any two-qubit pure state that is invariant under particle exchange up to a global phase shift $e^{i\varphi}$ trivially satisfies $P_{\text{and}} = P_{S_1} = P_{S_2} = P_{\text{or}}$ (see Appendix). More generally, for any two-qubit state $\rho$ that is invariant under particle exchange, $\rho=P\rho P$, where $P$ is the permutation matrix, the equality $P_{\rm and}=P_{S_1}=P_{S_2}=P_{\rm or}$ holds in every $m\times m$ dichotomic Bell scenario, including scenarios with marginal terms (see Appendix). Consequently, observing $P_{\rm or}>P_{\rm and}$ certifies that the underlying state is not permutationally invariant.

However, while all maximally entangled states share the same marginal violation
probability $P_S$, they do not all possess exchange invariance. This broken
symmetry manifests as a distinct difference between $P_{\text{and}}$ and
$P_{\text{or}}$. Consider the state
\begin{equation}
    \ket{\psi_\text{sym}} = \frac{1}{\sqrt{2}}
    \left(\ket{0}\ket{+}+\ket{1}\ket{-}\right),
\end{equation}
where $\ket{0}$, $\ket{1}$ and $\ket{+}$, $\ket{-}$ denote the eigenstates of
the Pauli matrices $\sigma_z$ and $\sigma_x$, respectively. This state is
maximally entangled and explicitly invariant under particle exchange, yielding
$P_{S_1} = P_{S_2} = P_{\text{and}} = P_{\text{or}} = 2(\pi-3)$.

Conversely, the maximally entangled state
\begin{equation}
    \ket{\psi_\text{asym}} = \frac{1}{\sqrt{2}}
    \left(\ket{0}\ket{-}+\ket{1}\ket{+}\right)
\end{equation}
breaks this symmetry, as is evident from the structure of its correlation
tensor,
\begin{equation}
    T_{\text{asym}} = \begin{pmatrix}
        0 & 0 & 1 \\
        0 & -1 & 0 \\
        -1 & 0 & 0
    \end{pmatrix},
\end{equation}
defined via $T_{ij} = \mathrm{Tr}[\rho\,(\sigma_i \otimes \sigma_j)]$.
For such an asymmetric state, $P_{\text{and}} \neq P_{S}$.
This divergence can be illustrated through a specific choice of observables.
Let Alice in $S_1$ measure along $A_1 = \sigma_y$ and $A_2 = \sigma_x$, and
Bob along $B_1 = (\sigma_z - \sigma_y)/\sqrt{2}$ and
$B_2 = -(\sigma_z + \sigma_y)/\sqrt{2}$.
In $S_2$, the settings are exchanged. Under these settings,
$\langle I_{S_1} \rangle = 2\sqrt{2}$ while $\langle I_{S_2} \rangle = 0$,
demonstrating that a maximal violation in $S_1$ is entirely compatible with
the absence of a violation in $S_2$. This is confirmed numerically: over $10^9$ randomly sampled measurement
settings we obtain
$P_{\text{and}} \approx 11.34\% \ll P_{S_1} \approx 28.32\%$.

The magnitude of this suppression depends strongly on the degree of state
asymmetry. As an example of a highly asymmetric maximally entangled state,
consider
\begin{equation}\label{asymmp}
    \ket{\psi_{\rm asym'}} = \frac{1}{\sqrt{2}}
    \Big(\ket{0}\ket{+} + i\ket{1}\ket{-}\Big),
\end{equation}
with correlation tensor
\begin{equation}
    T_{\rm asym'} = \begin{pmatrix}
        0 & -1 & 0 \\
        0 & 0 & 1 \\
        1 & 0 & 0
    \end{pmatrix}.
\end{equation}
Evaluating this state over $10^9$ randomly sampled measurements yields a
further suppressed joint violation probability of $P_{\text{and}} \approx
7.15\%$. Our numerical studies suggest that this is the minimal value of
$P_{\text{and}}$ among all maximally entangled two-qubit states.

The analysis above focused on specific maximally entangled states. A natural
extension is to consider random pure two-qubit states. As shown in
Table~\ref{tab:pure_states}, for $m = 2$ settings the violation probabilities
are significantly lower than those for maximally entangled states, but increase
strongly with the number of settings available to each observer.

\begin{table}[h]
    \centering
    \begin{tabular}{c|c|c|c|c|c} \hline \hline
         $m$ & $P_{S_1}$ [\%] & $P_{S_2}$ [\%] & $P_{\text{or}}$ [\%]
             & $P_{\text{and}}$ [\%] & Statistics \\ \hline
         2 & $5.3221$ & $5.3233$ & $9.2508$  & $1.3947$  & $10^9$ \\
         3 & $21.987$ & $21.985$ & $31.480$  & $12.492$  & $10^8$ \\
         5 & $50.016$ & $49.997$ & $59.226$ & $40.786$  & $10^7$ \\
         7 & $61.862$  & $61.714$  & $73.867$   & $49.709$   & $10^5$ \\
         \hline \hline
    \end{tabular}
    \caption{\label{tab:pure_states} Violation probabilities for random pure
    two-qubit states across varying numbers of measurement settings $m$ per
    observer.}
\end{table}

For completeness, we also evaluate the case of a random mixed state, sampled
$10^9$ times uniformly according to Hilbert-Schmidt distribution. For two
measurement settings we obtain $P_{S_1} \approx 0.000984\%$,
$P_{S_2} \approx 0.000986\%$, and $P_{\text{and}} \approx 0.000093\%$.

\section{Correlation measure}

\label{sec:correlation}
We now show that the asymmetry between $P_{\text{and}}$ and $P_S^2$ can be
used to quantify the statistical dependence between Alice's and Bob's
measurement randomness based solely on $P_{\text{and}}$. We begin with the
symmetric Bell state $\ket{\phi^+} = (\ket{00}+\ket{11})/\sqrt{2}$. As shown above, the exchange invariance of $\ket{\phi^+}$
guarantees $P_{\text{and}} = P_{S_1} = P_{S_2} = P_{S}$ when the measurement
settings in $S_2$ are obtained by directly swapping those of $S_1$. If instead
the settings in each experiment are drawn independently from a uniform random
distribution, the violations become statistically uncorrelated and the joint
probability factorizes as $P_{\text{and}} = P_{S}^2 = 4(\pi - 3)^2$. To capture this dependence quantitatively, we introduce the measure
\begin{equation}
    \Gamma = \frac{P_{\rm and} - P_S^2}{P_S (1-P_S)}.
    \label{gamma}
\end{equation}
For fully correlated measurement sets (direct swapping), $\Gamma = 1$; for
completely uncorrelated sets, $\Gamma = 0$. Intermediate regimes can be
modelled as probabilistic mixtures: with probability $p$ the parties use fully
correlated settings, and with probability $1-p$ they use independent ones,
yielding
\begin{equation}
    \Gamma = \frac{p P_S + (1-p) P_S^2 - P_S^2}{P_S(1-P_S)} = p.
\end{equation}
Thus $\Gamma$ interpolates smoothly between zero and one as the measurement
settings transition from fully independent to fully correlated.

\section{Finite measurement pools}

\label{sec:finite}
The inclusion-exclusion identity introduced in (\ref{ee}),
\begin{equation}
    P_{\text{or}} = P_{S_1} + P_{S_2} - P_{\text{and}},
\end{equation}
has a direct operational implication. Since $P_{S_1} = P_{S_2} = 2(\pi-3)$ is
the same for all maximally entangled states, a suppression of $P_{\text{and}}$
due to state asymmetry is necessarily accompanied by an enhancement of
$P_{\text{or}}$. This trade-off becomes practically significant when
measurement resources are limited.
The relevance of measurement settings as a resource for demonstrating nonlocality has been emphasized previously~\cite{Fonseca2015}. Here, we show that permutation asymmetry can further enhance the efficiency with which a finite set of settings reveals Bell nonlocality.

Consider a protocol in which Alice and Bob draw their measurement directions
from a restricted pool of $N = 4$ randomly sampled settings $D_1$, $D_2$, $D_3$ and $D_4$. They each select
two directions and attempt to violate local realism. If they do not observe the violation they return their settings to the pool, and repeat until all combinations are verified or violation has been observed. They obtain the probability of violating local realism with finite pool of $N$ measurement settings after averaging their result over many sampled pools.

For the symmetric state $\ket{\phi^+}$, one can construct 24 distinct
measurement sets from a pool of four directions $D_i$ eg. $\{D_2,D_3,D_1,D_4\}$, where Alice measures first two directions and Bob last two. Since we are accounting for violations of local realism, local swaps of measurements either on Alice's or Bob's side are meaningless thus reducing these 24 distinct
measurement sets to 6 tests, $\{D_1,D_2,D_3,D_4\}$, $\{D_1,D_3,D_2,D_4\}$, $\{D_1,D_4,D_3,D_2\}$, $\{D_3,D_2,D_1,D_4\}$, $\{D_4,D_2,D_3,D_1\}$, $\{D_4,D_3,D_2,D_1\}$. Additionally the exchange symmetry of $\ket{\phi^+}$ further halves the number of configurations to
3: $\{D_1,D_2,D_3,D_4\}$, $\{D_1,D_3,D_2,D_4\}$, $\{D_1,D_4,D_3,D_2\}$. The upper bound on the probability of observing at least one violation can be obtained assuming the 3 sets are uncorrelated. Then the resulting probability is equal $1-(1-2(\pi-3))^3=1-(7-2\pi)^3 \approx 63.17\%$,  where $(1-2(\pi-3))$ is the probability of not observing violation in a single test.  However, since the three sets are not uncorrelated the presented value is only an upper bound. To get the actual value for $\ket{\phi^+}$ state, we used Monte Carlo simulation, yielding the probability of approximately $53.97\%$.

Similarly for hypothetical fully asymmetric state, for which $P_{\text{and}} \to 0$, previously mentioned six configurations cannot be halved, because of broken permutation symmetry. Again obtaining upper bound using the same assumptions gives probability 
$1-(1-2(\pi-3))^6 \approx 86.43\%$. The value obtained in simulations for presented asymmetric state $\ket{\psi_{\text{asym'}}}$ (see Eq.~\eqref{asymmp}) for which $P_{\text{and}}\approx7.15\%$ reaches approximately $76.57\%$.
Asymmetric maximally entangled states 
therefore offer a concrete advantage in randomized Bell tests with finite
measurement pools.

For arbitrary size of the pool $N$, the upper bound on probability of not observing violation is equal to square root of upper bound on probability of not observing violation for the asymmetric state which is equal $(1-2(\pi-3))^{6\binom{N}{4}}$, where $\binom{N}{4}$ is the number of measurement sets sampled from finite $N$ element pool. The advantage of hypothetical fully asymmetric two qubit state for which $P_{\text{and}} \to 0$ over the symmetric two qubit state $\ket{\phi^+}$ can be quantified as ratio of the upper bounds on probabilities of violations obtained for these states
\begin{equation}
        \xi_A=\frac{1-(1-2(\pi-3))^{6\binom{N}{4}}}{1-(1-2(\pi-3))^{3\binom{N}{4}}}=1+(7-2\pi)^{3\binom{N}{4}}.
\end{equation}
From above we can see that the advantage is rapidly  approaching 1 so the advantage is the most prominent for small pools where for the above example of four element pool it is equal to  $(7-2 \pi )^3+1 \approx 1.36832$. 

\section{Conclusions}
\label{sec:discussion}

We have shown that permutation symmetry plays a decisive role in randomized Bell experiments with exchanged measurement settings. Although all maximally entangled two-qubit states exhibit the same probability of violating local realism under uniformly random measurements, they need not produce the same statistics across sequential Bell tests. In particular, we proved that every permutationally invariant bipartite state yields identical Bell-nonlocality outcomes in the original and swapped experiments, implying $ P_{\rm and} = P_{S_1} = P_{S_2} = P_{\rm or}, $
independently of the number of measurement settings and irrespective of the presence of marginal terms.

For states that are not invariant under particle exchange, this equivalence need not hold. The resulting suppression of the joint violation probability $P_{\rm and}$ reveals a reduced statistical dependence between the original and swapped experiments, an effect quantified by the correlation measure $\Gamma$ (\ref{gamma}). This observation further enables the detection of correlations between the measurement choices of Alice and Bob through the observed joint violation statistics.

The difference between symmetric and asymmetric states is not merely conceptual. Since a reduction of $P_{\rm and}$ necessarily increases $P_{\rm or}$, permutation asymmetry enhances the probability that at least one experiment reveals nonlocality. This effect becomes particularly relevant when measurement resources are limited. For a finite pool of randomly sampled measurement settings, we found that asymmetric maximally entangled states can substantially outperform their symmetric counterparts, achieving significantly higher probabilities of observing Bell nonlocality without enlarging the measurement pool.

More broadly, our results demonstrate that permutation asymmetry can become an operationally useful property even among states that are otherwise equivalent from the perspective of standard Bell-violation probabilities. Note that this perspective complements recent work on single Bell-inequality violations, where party-exchange asymmetry can also be useful, but where the relevant asymmetry lies in the Alice-Bob measurement strategy rather than necessarily in the shared state~\cite{Hsu2026}. By contrast, our results show that permutation asymmetry of the shared state itself can become an operational resource in randomized Bell tests, despite being invisible at the level of single-test violation probabilities, thereby motivating further investigations of asymmetry-based advantages in multipartite systems, higher-dimensional quantum states, and device-independent quantum-information protocols.

\section{Acknowlegments} 

We thank Marcin Paw{\l}owski for fruitful discussions. WL and JW are supported by the National Science Centre (NCN, Poland) within the OPUS project (Grant No. 2024/53/B/ST2/04103). TV acknowledges support from the European Union (CHIST-ERA MoDIC) and from the National Research, Development and Innovation Office NKFIH (Grant Nos.~2023-1.2.1-ERA\_NET-2023-00009 and K145927).

\appendix

\section{Permutation symmetry and invariant states}


In this appendix we characterize pure two-qubit states for which sequential Bell experiments with swapped observables yield the same violation statistics regardless of the particular measurement choices. Suppose
the state $\ket{\psi}$ is invariant under a combined permutation and local
unitary operation,
\begin{equation}
\label{psiinv}
    \ket{\psi} = (C \otimes I)\,P\ket{\psi},
\end{equation}
where $C$ is a general unitary and $P$ is the permutation matrix
\begin{equation}
\label{Pmat}
    P = \begin{pmatrix}
        1 & 0 & 0 & 0 \\
        0 & 0 & 1 & 0 \\
        0 & 1 & 0 & 0 \\
        0 & 0 & 0 & 1
    \end{pmatrix}.
\end{equation}
We quantify the difference between outcomes of the two swapped experiments as
\begin{eqnarray}
    \delta &=& I_{S_1} - I_{S_2} \nonumber \\
    &=& \langle A_1 B_1 \rangle + \langle A_2 B_1 \rangle
      + \langle A_1 B_2 \rangle - \langle A_2 B_2\rangle \nonumber \\
    &&- \bigl(\langle B_1 A_1 \rangle + \langle B_1 A_2 \rangle
      + \langle B_2 A_1\rangle - \langle B_2 A_2 \rangle\bigr).
\end{eqnarray}
Using the invariance condition, one can show that
\begin{eqnarray}
    \bra{\psi} A \otimes B \ket{\psi}
    &=& \bra{\psi}\bigl(P(C^\dagger \otimes I)\bigr)
        (A \otimes B)\bigl((C \otimes I)P\bigr)\ket{\psi} \nonumber \\
    &=& \bra{\psi} P\,(C^\dagger A C \otimes B)\,P \ket{\psi} \nonumber \\
    &=& \bra{\psi} B \otimes C^\dagger A C \ket{\psi} \nonumber \\
    &=& \bra{\psi} B \otimes A \ket{\psi}
      + \bra{\psi} B \otimes C^\dagger[A,C] \ket{\psi}.
\end{eqnarray}
Substituting into $\delta$ yields
\begin{eqnarray}
    \delta &=& -\langle A_1 \otimes C^\dagger[B_1,C] \rangle
             - \langle A_2 \otimes C^\dagger[B_1,C] \rangle \nonumber \\
           &&- \langle A_1 \otimes C^\dagger[B_2,C] \rangle
             + \langle A_2 \otimes C^\dagger[B_2,C] \rangle.
\end{eqnarray}
This expression vanishes for all measurement settings if and only if $C$ is proportional to the identity, $C=e^{i\varphi}I$. Substituting this form into Eq.~\eqref{psiinv} gives $P\ket{\psi}=e^{-i\varphi}\ket{\psi}$. Since $P^2=I$, the corresponding swap eigenvalue must be $\pm1$. Thus, within the pure-state condition of Eq.~\eqref{psiinv}, the relevant states belong precisely to the symmetric and antisymmetric subspaces, $P\ket{\psi}=\pm\ket{\psi}$. These are the states which produce identical violation statistics in the sequential Bell tests with swapped measurement settings, as claimed in the main text.\\

\section{Swapped Bell tests with invariant states}

In this appendix we show that, in any bipartite $m$-setting, two-outcome Bell scenario with marginal terms included, a permutationally invariant state gives identical probability of violation in the original and swapped experiments. That is, $P_{\rm and}=P_{S_1}=P_{S_2}=P_{\rm or}$.

Let Alice and Bob measure $m$ dichotomic observables $A_x$ and $B_y$ for all $x,y=1,\ldots,m$. Also, introduce the auxiliary observables $A_0=B_0=I_d$, where $d$ is the local state space dimension. The corresponding correlation matrix is
\begin{equation}
    E_{xy}
    =
    \text{Tr}\left[\rho(A_x\otimes B_y)\right],
    \qquad x,y=0,1,\ldots,m.
\end{equation}
Thus $E_{00}=1$, the entries $E_{x0}$ are Alice's marginals, the entries $E_{0y}$ are Bob's marginals, and the entries $E_{xy}$ with $x,y\geq 1$ are the usual two-body correlators.

Let $P$ denote the flip operator, which for arbitrary local operators $X$ and $Y$ works as
\begin{equation}
    P(X\otimes Y)P=Y\otimes X.
\end{equation}
On the other hand, it reduces to Eq.~\eqref{Pmat} for $d=2$.

Assume that the bipartite state is permutationally invariant, $\rho=P\rho P$. In the swapped experiment, Alice measures $B_x$ and Bob measures $A_y$. The corresponding correlation matrix is therefore
\begin{equation}
    E^F_{xy}
    =
    \text{Tr}\left[\rho(B_x\otimes A_y)\right] 
    =
    \text{Tr}\left[\rho(A_y\otimes B_x)\right]  
    =
    E_{yx}.
\end{equation}
Hence, $E^F= E^T$.

To complete the proof, note that transposition is a symmetry of the full local $m$-setting Bell polytope, including marginal terms. Indeed, deterministic local vertices have the form
\begin{equation}
    E_{xy}=a_x b_y,
    \qquad
    a_0=b_0=1,
    \qquad
    a_x,b_y\in\{\pm1\}.
\end{equation}
Equivalently, $E=a b^T$. Under transposition, $E^T=b a^T$,
which is again a deterministic local vertex, with Alice's and Bob's deterministic strategies exchanged. Since the local Bell polytope is the convex hull of these deterministic vertices, transposition maps the polytope onto itself. Therefore, $E$ is local if and only if $E^T$ is local.

Thus, for every fixed choice of measurement settings, the original and swapped experiments either both admit a local model or both violate the local Bell polytope. The actual Bell violations in $S_1$ and $S_2$ therefore coincide for any fixed measurement settings. Consequently,
\begin{equation}
    P_{\rm and} = P_{S_1 \wedge S_2}
    = P_{S_1} = P_{S_2}
    = P_{S_1 \vee S_2} = P_{\rm or}.
\end{equation}


\bibliographystyle{apsrev4-2}
\bibliography{ref}

\end{document}